\documentclass{aa}
\usepackage{psfig}

\def\ks{K$_{\rm s}$}
\def\jmk{J$-$K}
\def\jh{$_{\rm jh}$}
\def\hk{$_{\rm hk}$}

\def\chisq{$\chi^2$}
\def\rms{$\sigma_m$}
\def\absdev{$\overline{|m_d|}$}

\def\micron{$\mu$m}
\def\kms{kms$^{-1}$}

\def\h2o{H$_{\rm 2}$O}
\def\c02{CO$_{\rm 2}$}
\def\teff{${\rm T}_{\rm eff}$}
\def\logg{log\,g}
\def\msol{M$_{\odot}$}

\begin{document}

\title{Dust clouds or magnetic spots? Exploring the atmospheres
of L~dwarfs with time-resolved spectrophotometry}
\subtitle{} 
\titlerunning{atmospheric variability of ultra cool dwarfs}

\author{C.A.L.\ Bailer-Jones}
\authorrunning{C.A.L.\ Bailer-Jones}
\institute{Max-Planck-Institut f\"ur Astronomie, K\"onigstuhl 17, D-69117 Heidelberg, Germany, calj@mpia-hd.mpg.de}

\date{Received 27 March 2002 / Accepted 15 May 2002}

\abstract{I present the results of a program to spectrophotometrically
monitor the L1.5 dwarf \object{2MASSW J1145572+231730} to identify the
cause of photospheric variability in ultra cool dwarfs. Plausible
candidates are magnetically-induced star spots and inhomogeneous
photospheric dust clouds. Based on the atmospheric models and
synthetic spectra of Allard et al.\ (2001), the expected signatures of
these phenomena in the 0.5--2.5\,\micron\ wavelength region are
presented and discussed.  Near infrared spectra of 2M1145 were
obtained along with a nearby reference star observed simultaneously in
the spectrograph slit.  No convincing variability over a 54 hour
interval is found in any one of several colour indices designed to be
most sensitive to dust- and spot-related variability. Nonetheless, a
significant {\it correlation} between the variability of two colour
indices is found.  This is slightly more consistent with the
dust-related variability model than the cool spot one considered.
Based on the theoretically predicted signatures and the median errors
in the colour indices (0.03-0.05 magnitudes), upper limits are placed
on the coverage of possible spots and clouds. Assuming the L dwarf to
be best modelled by a dusty atmosphere at 1900\,K, coherent clear
clouds are limited to a coverage of 10--15\% of the projected surface
area and 200\,K cooler spots to a 20\% coverage.  A larger coverage of
many small features varying incoherently cannot be ruled out with this
method.  A lower effective temperature restricts coherent clear clouds
to be much smaller; a higher temperature allows both clouds and spots
to be larger.  These upper limits are consistent with the two separate
variability {\it detections} in the I-band reported by Bailer-Jones \&
Mundt (2001).


}
\maketitle

\section{Introduction: variability, dust and magnetic fields in ultra cool dwarfs}
\label{intro}

In a study of 21 low mass stars and brown dwarfs, Bailer-Jones \&
Mundt (\cite{bjm}; hereafter BJM) found evidence for I-band
photometric variability in 11 M and L dwarfs, with variability
timescales of a few hours and RMS amplitudes of 0.01--0.06\,mags. In a
few cases, this variability could not be ascribed to a rotational
modulation by stable surface features (e.g.\ cool spots), such as
those often found on pre-main-sequence, earlier-type stars (e.g.\
Herbst, Bailer-Jones \& Mundt \cite{herbst01}).  This observed
non-periodic variability was therefore interpreted as evidence for the
evolution of surface features on a timescale shorter than a rotation
period.  This could happen if the features rapidly form and dissipate
or if they migrate with respect to the rotation.

This raised the question as to the physical nature of the putative
surface features on ultra cool dwarfs (UCDs).\footnote{An ultra cool
dwarf (UCD) is loosely defined here as any dwarf star with spectral type
later than about M8, thus including L and T dwarfs.  UCDs have
effective temperatures below about 2800\,K, surface gravities in the
range 300--3000\,ms$^{-2}$ (i.e.\ \logg\ of 4.5--5.5) and masses less
than about 0.1\,\msol. Following the mass definition of substellar
objects, a given UCD could be a low mass hydrogen-burning star, a
brown dwarf or a gas giant planet.}  BJM found no correlation between
the presence or amplitude of variability and the H$\alpha$ equivalent
width (although many of the objects had no H$\alpha$ data). Taking the
latter as a measure of chromospheric activity and hence as a proxy for
the size and/or temperature contrast of magnetically-induced
photospheric spots, this lack of correlation provides evidence against
the surface features being magnetic spots.  This is supported by the
observation of Gizis et al.\ (\cite{gizis00}), who found that the
ratio of H$\alpha$ emission to bolometric luminosity decreased by more
than two orders of magnitude from M7 to L1.  As BJM also found that
variability appeared to be more common later than M9, they
concluded that their observed variability was probably not magnetic in
nature. 

It has now been well established that dust grains form in the
atmospheres of UCDs, and that at least some component of these remain
suspended in the photosphere at effective temperatures above about
1800\,K (Burrows \& Sharp \cite{burrows99}; Lodders \cite{lodders99};
Chabrier et al.\ \cite{chabrier00}; Allard et al.\ \cite{allard01};
Schweitzer et al.\ \cite{schweitzer01}; Marley et al.\
\cite{marley02}). The presence of solid condensates combined with the
fact that UCDs are probably fully convective (Chabrier \& Baraffe
\cite{chabrier00b}) and are often rapid rotators (Basri et al.\
\cite{basri00}), provides a mechanism for turbulent processes to drive
dust dynamics and produce time-dependent opacities. BJM took this to
be the single most likely mechanism to explain their observed
variability.

However, doubt about the general absence of magnetic activity in L
dwarfs has been raised by the detection of radio emission in four UCDs
(of a sample of 13) by Berger (\cite{berger02}) and Berger et al.\
(\cite{berger01}).  From the intensity of the continuous emission and
flaring at 8.5\,GHz detected in these objects, Berger
(\cite{berger02}) concludes that the emission mechanism is most likely
synchrotron and not thermal, indicating the presence of magnetic
fields and electron column densities similar to those inferred for
earlier-type flaring M dwarfs. Although based on a very small number
of detections, these data also hint that the radio emission is no
longer correlated with the H$\alpha$ emission in the M7--L1 spectral
type region where the latter decreases sharply, the implication being
that H$\alpha$ emission no longer traces magnetic activity.

While magnetic fields may well be present in UCDs, this does not
necessarily mean that they produce photospheric spots.  UCDs have very
cool atmospheres and low ionization fractions (e.g.\ Meyer \&
Meyer-Hofmeister \cite{meyer99}; Fleming et al.\
\cite{fleming00}). Thus there are relatively few free electrons and so
little coupling between the gas and any magnetic field, with the
result that cool spots may not be able to form in the visible
photosphere.  Gelino et al.\ (\cite{gelino02}) have attempted to
quantify this lack of coupling by showing that the magnetic Reynolds
number is probably very low\footnote{However, spots could still form
deeper in the atmosphere where the Reynolds number is higher due to
the lower ionization, and these may be visible if the atmosphere is
less opaque due to the precipitation of dust (see section
\ref{models}).} ($<1$) for UCDs with convective velocities below 100
ms$^{-1}$.  This neutrality of UCD photospheres can be reconciled with
the relatively high electron column densities inferred by Berger
(\cite{berger02}) on the assumption that the radio emission is
primarily due to a high density of {\it coronal} electrons. In their
own I-band monitoring of 18 L dwarfs, Gelino et al.\ (\cite{gelino02})
found seven to be variable, and concluded on the basis of the
neutrality of L dwarfs that the variability was not magnetic in
origin. Further, Mart\'\i n et al.\ (\cite{martin01}) detected
photometric variability in the field M9 dwarf BRI\,0021$-$0214, which
generally shows very weak magnetic activity, although an H$\alpha$
flare has been observed from this object (Reid et al.\
\cite{reid99}).

This returns us to the behaviour of dust.  In a cool static
atmosphere, dust grains will form, grow and gravitationally settle
below the photosphere creating a dust-cleared photosphere depleted in
those elements locked into the grains.  By contrast, in a dynamic
atmosphere, convection will provide a bouyancy acting against
precipitation, and deep convection may recycle precipitated elements
back into the visible atmosphere, thus retaining a dusty composition.
First attempts to consider the relative efficiency of dust settling
and convective recycling -- and hence the {\it vertical} distribution
of dust grains in the photosphere -- have been made by Tsuji
(\cite{tsuji01}) and Ackerman \& Marley (\cite{ackerman01}). A gradual
transition from a dusty to a dust-cleared atmosphere is predicted to
occur with decreasing effective temperature.  Dust can be considered
to be confined to a thin layer, the depth (pressure) of the base of
which is determined by the condensation temperature of grains.
Condensation can occur throughout the atmosphere above this base with
increasing efficiency at higher altitudes (due to lower temperatures),
although with decreasing contribution to the opacity on account of the
decreasing density. A limit on the vertical extent of this layer may
be set by the growth of grains above some critical size, at which
point gravity overcomes turbulent support and these grains are
precipitated out into deeper layers of the atmosphere. As the
effective temperature decreases, the base of this dust layer moves to
deeper parts of the atmosphere, where the integrated gas opacity (as
seen from outside the photosphere) is increased. Hence the dust layer
becomes less visible with the result that the spectrum increasingly
represents a dust-cleared atmosphere.  Such a mechanism may explain the
switch from red to blue infrared colours for the transition from mid L
dwarfs to T dwarfs.

Neither the model of Tsuji (\cite{tsuji01}) nor that of Ackerman \&
Marley (\cite{ackerman01}) addresses the possibility of {\it
horizontal} inhomogeneities or dust grain dynamics which would be
required to produce  variability in the unresolved disk (but
see Helling et al.\ \cite{helling01}). In a dusty or dust-cleared
atmosphere, these inhomogeneities could be regions of enhanced dust
formation and  dust opacity, perhaps due to a localized deep
convective column bringing dust forming material up from the deeper,
opaque layers of the atmosphere. Alternatively, in a dusty atmosphere,
dust formation may be slow in the material brought up by deep
convection, thus creating a short-lived dust-free cloud.

\section{Current objectives and target information}
\label{objectives}

In the present paper, I report the results of an observational program
to investigate further the physical cause of variability in UCDs. Such
an investigation is naturally rather speculative given how little we
known of time-dependent atmospheric processes in UCDs. In light of
this, I decided to spectrophotometrically monitor a UCD over a wide
wavelength range. The near infrared (NIR) region (0.9--2.5\,\micron)
was chosen over the optical (0.3--1.0\,\micron) because UCDs are
significantly brighter there.

At the time of target selection (July 1999), the only known
photometrically variable UCDs were the L1.5 dwarf \object{2MASSW
J1145572+231730} (Bailer-Jones \& Mundt \cite{bjm99}) and the M9 dwarf
LP944-20 (Tinney \& Tolley \cite{tinney99}).  The former was selected
because (1) we intended to monitor it again in the I-band and (2) it
has a nearby bright reference star which can be used to obtain {\it
differential} spectrophotometry. Due to rapid changes in sky
transmission and brightness in the NIR, it was felt that
spectrophotometry from regular calibration against standards was not
viable. Instead, spectrophotometry is done relative to a second star
in the same slit, on the assumption that this reference star is
photometrically stable to some required level on the timescale of
interest.  This reference star (which is not presently in NED or
SIMBAD and is saturated in the data of BJM) has designation
\object{2MASSI J1145560+231613}, and has B,R,J,H and \ks\ magnitudes
of 14.80, 14.10, 13.57, 13.28 and 13.18 respectively, as taken from
the 2MASS second incremental release catalogue.  These are most
consistent with an unreddened or slightly reddened mid- or late-type F
dwarf. The Galactic latitude of the field is $l=226^{\circ}$,
$b=75^{\circ}$, implying that it is probably an older thick disk or
halo star at 1-2\,kpc, rather than a very young potentially variable
star or pulsating supergiant.

\object{2MASSW J1145572+231730} (hereafter 2M1145) is an L1.5 dwarf
discovered by 2MASS with I,J,H and \ks\ magnitudes of 18.6, 15.37,
14.52 and 13.92 respectively (Kirkpatrick et al.\
\cite{kirkpatrick99}; the J,H,\ks\ magnitudes are from the 2MASS
second incremental release catalogue). The optical spectrum of
Kirkpatrick et al.\ (\cite{kirkpatrick99}) shows an H$\alpha$ emission
line with an equivalent width of 4.2\,\AA.\footnote{The statement on
p.\ 507 of Basri (\cite{basri00b}) that 2M1145 has shown flaring in
H$\alpha$ is incorrect. The paper referenced cites flaring in a
different object.}  The absence of strong lithium absorption implies
that it may be a low mass star rather than a brown dwarf, but its late
spectral type and apparent relative youth ($<$\,1\,Gyr) mean that it
probably has a mass close to the stellar/substellar boundary at
0.075--0.08\,\msol.

Based on data taken in January 1999, Bailer-Jones \& Mundt
(\cite{bjm99}) found 2M1145 to be photometrically variable in the
I-band with an RMS amplitude of 0.038 mags and a tentative period of
7.1 hours. An improved analysis of these data confirmed the detection
of a significant variability -- albeit with a slightly smaller
amplitude (0.031 mags) -- but did not confirm the period (BJM). New
data taken simultaneously with that in the present paper (February
2000), and reported in BJM, made a much more significant detection of
variability (amplitude 0.020 mags), again over a timescale of a few
hours. The nature of the power spectra, plus the significant
differences between the 1999 and 2000 power spectra, led BJM to
conclude that the surface features evolved over a timescale of a few
hours (see section \ref{intro}). Since then, variability has been
discovered in a number of other late M and early to mid L dwarfs, some
also showing non-periodic variability (Mart\'\i n et al.\
\cite{martin01}; Clarke et al.\ \cite{clarke02}; Gelino et al.\
\cite{gelino02}) and water variability has been claimed in a T dwarf
(Nakajima et al.\ \cite{nakajima00}).

The layout of the rest of this paper is as follows. In the next
section I describe the observational signatures of different types of
UCD atmospheric variability, based on the atmospheric models and
synthetic spectra of Allard et al.\ (\cite{allard01}).  In then
describe the observational procedure used (section \ref{data})
followed by the method of spectral extraction and combination (section
\ref{processing}).  The variability analysis is presented in section
\ref{analysis} and the implications of these results for atmospheric
changes in UCDs is given in section \ref{discussion}.  While no
convincing variability signal was detected, the data nonetheless
permit various extreme cases to be ruled out. The main points of the
paper are summarised in section \ref{conclusions}.

\section{Observability of theoretical predictions of atmospheric features}
\label{predictions}

\subsection{Method and models}
\label{models}

Predictions of the photometric signatures of surface
features on UCDs can be obtained from combining synthetic spectra
which represent (a) the ``background'' (feature-free) atmosphere and
(b) the feature, in proportion to the surface coverage of the feature.
For this purpose I will use the AMES-dusty and AMES-cond models of
Allard et al.\ (\cite{allard01}). Both models permit dust grains to
form, but they represent the extreme cases of dust precipitation.  In
the AMES-dusty model (hereafter {\it dusty}), precipitation is
prevented (e.g.\ convective recycling is very efficient), whereas in the
AMES-cond model (hereafter {\it cond}) the gravitational settling of
dust efficiently clears the atmosphere of all condensed dust (and
of course its constituents).  In reality, intermediate situations probably
exist with varying degrees of dust precipitation (e.g.\ Tsuji
\cite{tsuji01}; Ackerman \& Marley \cite{ackerman01}). Therefore, the
following predictions serve to examine the extreme effects which may
occur and which can be observationally tested.

Both the dusty and cond models of Allard et al.\ (\cite{allard01})
assume that spherical dust grains form in chemical equilibrium with
the gas phase.  In the cond model this dust is removed from the
atmosphere. The authors assume an interstellar size distribution of
dust grains between 0.006 and 0.240\,\micron.  They acknowledge that
this is somewhat arbitrary, but provided that the largest particles
are sufficiently small such that scattering remains in the Rayleigh
regime for the wavelengths considered ($\lambda > 0.5$\,\micron), this
is of no relevance to the formation of the spectral energy
distribution.  Convection is treated using the mixing length theory
with a mixing length equal to the pressure scale height and a
microturbulence velocity of 2\,\kms. While convection is included
for calculating the energy transport, it is not included in the
determination of the dust number densities. So in these models,
convection is not a mechanism for returning precipitated dust elements
back to the photosphere.

In the following, I consider a cloud on a dusty atmosphere to be a
compact region cleared of dust (i.e.\ represented by the cond model)
in thermal equilibrium with the ``background'' atmosphere, that is, at
the same effective temperature. Clearly this assumption would not hold
if material is brought up rapidly from deep within the star to form
the cloud.  Likewise, a cloud on a cond atmosphere is represented by
the dusty model, again at the same effective temperature as the
background atmosphere.  Magnetic spots are modelled as compact regions
with the same atmosphere (dusty or cond) as the background atmosphere,
but with a lower effective temperature.

Of course, other causes of photometric variability in UCDs may
exist. Newly formed UCDs may have accretion disks, infall of matter
onto which could create observable hot spots, as could interactions
with a close companion. However, magnetic fields, dust and atmospheric
dynamics are ubiquitous, so a search for the observational signatures
of these seems a sensible first step.

\subsection{Cloud and spot signatures at 1900\,K}

\begin{figure}[t]
\centerline{
\psfig{figure=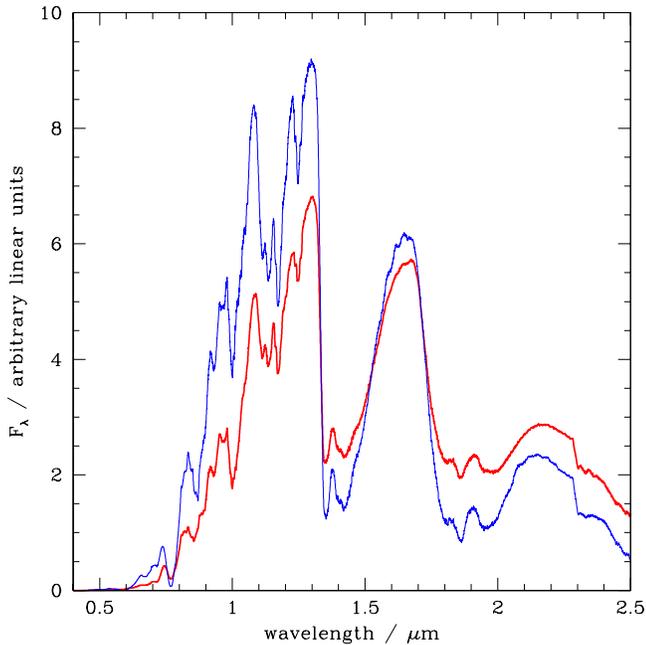,angle=0,width=0.5\textwidth}
}
\caption{AMES-dusty (thick/red line) and AMES-cond (thin/blue line)
synthetic spectra from Allard et al.\ (2001) for \teff\,=\,1900\,K and
\logg\,=\,5.5, the nominal parameters for the L1.5 field dwarf
2M1145. These spectra, as well as those in Figs.\
\ref{var5}--\ref{var6}, are shown at a resolution of
0.01\,\micron. [Colour]}
\label{sed1}
\end{figure}

Based on the fits of the Allard et al.\ (\cite{allard01}) synthetic
spectra to low and high resolution observations of L dwarfs by
Schweitzer et al.\ (2001), an L1.5 spectrum (the spectral type of
2M1145) is most consistent with a dusty model with \teff\,=1900\,K and
\logg\,=\,5.5, although models with $\pm 100$\,K and $\pm 0.5$\,dex in
\logg\ are equally valid depending on which data are used.
Fig. \ref{sed1} shows this model spectrum from Allard et al.\ (2001),
plus the cond spectrum at the same temperature and gravity for
comparison.  Dusty models explain well the NIR colours of UCDs down to
mid L dwarfs (circa 1800\,K) (e.g.\ Chabrier et al.\ \cite{chabrier00}).
At \teff\,=\,1900\,K, dust grains form deep in the photosphere with
the result that the photosphere is heated by several hundred Kelvin
and the whole photosphere is moved `up' in the atmosphere, i.e.\ to
lower gas pressures (cf.\ Figs.\ 5 and 6 of Allard et al.\
\cite{allard01}).  This dust opacity suppresses the flux blueward of
about 1.3\,\micron\ and leads to it being re-radiated at longer
wavelengths. At these longer wavelengths, water steam is the dominant
source of opacity, as can be seen from the deep water absorption bands
at 1.35--1.45 and 1.8--2.0\,\micron\ in Fig.\ \ref{sed1}. The cond
model atmospheres, by contrast, are more transparent and cooler, due
to the lack of both dust grains and the molecular species which were
locked into the dust.

\begin{figure}[t]
\centerline{
\psfig{figure=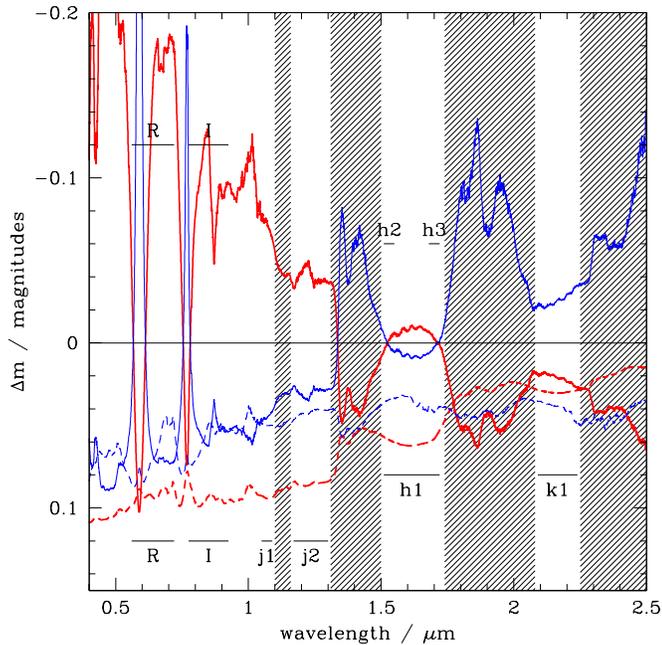,angle=0,width=0.5\textwidth}
}
\caption{The change in spectrum of an ultra cool dwarf with
\teff\,=\,1900\,K and \logg\,=\,5.5 due to the formation of a cloud or
cool spot with a coverage factor of 0.1 according to equation
\ref{sfeq}.  The four lines shown are for the following cases: cond
cloud on a dusty atmosphere (thick/red solid line) and 200\,K cooler
spot on a dusty atmosphere (thick/red dashed line); dusty cloud on a
cond atmosphere (thin/blue solid line) and 200\,K cooler spot on a
cond atmosphere (thin/blue dashed line).  The cool spot is modelled
using the same atmosphere type (cond or dusty) as the ``background''
(feature-free) atmosphere.  The grey bands show regions of telluric
absorption, i.e.\ non- or poorly accessible from ground-based
observations. The horizontal labelled lines show the location of
various bands defined in Table \ref{bands}. The FWHM of the R and
I-bands are also shown (the latter being that used by Bailer-Jones \&
Mundt 1999,2001). [Colour]}
\label{var5}
\end{figure}

Fig. \ref{var5} shows the change in a 1900\,K dusty spectrum
due to the formation of a cond cloud covering 10\% of the surface
(thick/red solid line). It is shown in terms of the conventional magnitude
system, such that the ordinate is

\begin{equation}
\Delta m(\lambda) = -2.5 \log_{10}\left(\frac{eF_c(\lambda) + (1-e)F_a(\lambda)}{F_a(\lambda)} \right)
\label{sfeq}
\end{equation}

\noindent where $F_a(\lambda)$ is the model spectrum
(Wm$^{-2}$\AA$^{-1}$) of the cloud-free (background) atmosphere,
$F_c(\lambda)$ the spectrum of the cloud, and $e$ the cloud covering
fraction, in this case 0.1. Fig. \ref{var5} also shows this for a
dusty cloud on a cond atmosphere (thin/blue solid line).  Thus during a
period of cloud formation and dissipation, the spectrum is predicted
to change by the amount shown.

The formation of a cond cloud on a dusty atmosphere increases the flux
over almost all of the region 0.5--1.3\,\micron\ due to the reduced
total dust opacity in the integrated disk in this wavelength
region. Likewise, the lower temperature of the cond cloud results
in the recombination of more water molecules and thus a larger opacity
at wavelengths above 1.3\,\micron.

Another consequence is an increase in the strength of the
absorption lines of the neutral alkali elements (Li,Na,K,Rb,Cs) . This
occurs because the lower temperature cond cloud decreases the opacity
below these absorbers, thus increasing the contrast of these
lines. This effect is most visible (see Fig.\ \ref{var5}) in the
highly broadened Na{\small I} doublet (0.59\,\micron) and K{\small I}
doublet (0.77\,\micron), the broadening also a consequence of the
higher transparency of the cond cloud such that the increased pressure
broadening of the lines deeper in the photosphere is seen.

The spectral changes due to a dusty cloud on a cond atmosphere at
\teff\,=\,1900\,K are of the opposite sign and on a smaller scale, but
otherwise have a very similar wavelength dependence.  Because, when we
observe a star, we do not know a priori the phase of the variations,
i.e.\ when the star is cloud-free, spectrophotometry alone will not
distinguish between these two cases: {\it The formation of one feature
looks like the dissipation of the other}. For example, the formation
of a cond cloud on a dusty atmosphere will make the star get brighter
in J as it gets bluer in \jmk, but this is the same signature as the
dissipation of a dusty cloud on a cond atmosphere (Fig.\ \ref{var5}).
The amplitude of the signatures differs for a given cloud coverage,
but this too is unknown a priori.  Only very accurate measurements
could distinguish between the small differences in the wavelength
dependencies for these two cases. If we could assume that surface
features are rare and short-lived, the cloud-free atmosphere could be
determined over time. However, the fraction of variable UCDs
discovered belies this option (section \ref{intro}).

\subsection{Photometric signatures in flux integrated bands}
\label{bandsec}

Rather than examining the variability at all resolved wavelengths in
the data presented later, I have chosen to define various bands by
summing the flux over continuous wavelengths regions where the
variability signature is similar (Table \ref{bands}). The bands j1,
j2, h1 and k1 were chosen to avoid regions of strong telluric
absorption.  Variability of water absorption in the Earth's atmosphere
makes reliable detection in such regions very difficult.  h1 and k1
are essentially the conventional H and K bands shortened to minimize
sensitivity to this. The blue end of j1 was set to avoid the region
1.00--1.05\,\micron, which turned out to be very noisy in the spectra
presented later in this paper.  h2 and h3 show essentially no flux
variability under cloud formation whereas they do vary under spot
formation.  These two bands were set specifically for a UCD with
\teff\,$\simeq$\,1900\,K, i.e.\ 2M1145, but as can be seen from
Figs. \ref{var3} and \ref{var4}, they still have only a small
signature at both higher and lower temperatures.

\begin{table}[t]
\caption[]{Wavelength bands for analysing near infrared spectra for
variability effects. The subscripts jh and hk are used to distinguish
whether the H bands were measured from the JH or HK spectra (see
section \ref{data}).}
\label{bands}
\begin{tabular}{ll}\hline
band	& Wavelength range / \micron \\
\hline
j1	& 1.05--1.09	\\
j2	& 1.17--1.30	\\
h1	& 1.51--1.72	\\
h2	& 1.51--1.55	\\
h3	& 1.68--1.72	\\
k1	& 2.09--2.24	\\
\hline
\end{tabular}
\end{table}

The most useful discriminant between a cond cloud and a cool spot on a
dusty atmosphere comes from the j1 and k1 bands
(Fig. \ref{var5}). With the cloud covering 10\% of the surface, the
change in j1 is {\it anticorrelated} with that in k1, with amplitudes
of 0.07\,mags and 0.025\,mags respectively. In contrast, with a cool
spot, j1 and k1 are {\it correlated} with amplitudes of 0.09\,mags and
0.03\,mags respectively. This distinguishes between these scenarios
without prior knowledge of the cloud/spot size or variability phase
(i.e.\ the band fluxes in the absence of surface features).  

Unique signatures also exist in the colours formed from these bands.
These can serve to confirm a putative model from the same spectral
data. Further, in some circumstances, only colour and not absolute
photometric data may be available.  The formation and dissipation of a
cond cloud causes the colours j1$-$j2, j1$-$h2, j2$-$h2 and h2$-$k1 to
vary (in the same direction) with amplitudes of 0.03, 0.07, 0.04 and
0.025\,mags respectively. The cool spot, on the other hand, produces
variations (again in phase) in these colours of 0.005, 0.035, 0.03 and
0.03 mags respectively. Although all these colour variations are in
the same direction for a given model, the fact that their amplitudes
differ enable us to distinguish between the models without having to
know a priori the spot size. The amplitudes scale linearly for small
feature coverages so the amplitude ratios permit a distinction.  For
the cond cloud, j1$-$j2, j1$-$h2, j2$-$h2 and h2$-$k1 have amplitude
ratios of 3:7:4:3, whereas for the cool spots this is 1:7:6:6.  The
band and colour signatures are summarised in Table \ref{signatures}.

\begin{table}[t]
\caption[]{The predicted photometric variations (in magnitudes) due to
the formation of various types of surface features covering 10\% of
the surface of an ultra cool dwarf with \teff\,=\,1900\,K and
\logg\,=\,5.5, the nominal parameters for 2M1145. The signatures are
shown for the two types of atmosphere and two types of features
plotted in Fig.\ \ref{var5}: the clouds have the same effective
temperature as the atmosphere; the spots are 200\,K cooler. In the
usual convention, positive signs denote decreased brightness (bands)
or increased redness (colours).}
\label{signatures}
\begin{tabular}{lrrrr}\hline
  band or   & \multicolumn{2}{c}{dusty atmosphere} & \multicolumn{2}{c}{cond atmosphere} \\
  colour        & cond       & cool       & dusty       & cool               \\
                & cloud      & spot       & cloud       & spot               \\
\hline
j1              & $-$0.070   & 0.090      & 0.045       & 0.050              \\
j2              & $-$0.040   & 0.085      & 0.030       & 0.045              \\
h1              & $-$0.005   & 0.060      & 0.005       & 0.035              \\
h2              & 0.000      & 0.055      & 0.000       & 0.035              \\
h3              & 0.000      & 0.055      & 0.000       & 0.040              \\
k1              & 0.025      & 0.030      & $-$0.030    & 0.040              \\
j1$-$j2         & $-$0.030   & 0.005      & 0.015       & 0.005              \\
j1$-$h2         & $-$0.070   & 0.035      & 0.045       & 0.015              \\
j2$-$h2         & $-$0.040   & 0.030      & 0.030       & 0.010              \\
h2$-$k1         & $-$0.025   & 0.030      & 0.030       & $-$0.005           \\
\hline
\end{tabular}
\end{table}

In order to be able to discriminate between these various scenarios at
the 3$\sigma$ level for a 10\% feature coverage, these amplitudes
would have to be measured to an accuracy of at least 0.01\,mags. In
principle, these amplitudes can be used to derive the spot size,
although this calls for even higher accuracy.  How these required
accuracies translate to individual photometric accuracies depends on
the number of measurements used to determine the amplitude: the
individual measurements do not need to obtain this level of
photometric precision in their random errors. However, the most
challenging problem will be to keep the systematic errors\footnote{By
{\it systematic error} I mean any error which can make the measured
value (e.g.\ the flux) correlate with some observing parameter, such
as time, airmass, wavelength or position on the detector. A {\it
random error}, by contrast, is one in which the measured value shows
no such correlation.} well below 1\%.  Possible significant effects
are imperfect background subtraction (especially in the NIR) and
colour-dependent flux losses due to decentering of the star on the
spectrograph slit (see sections \ref{data} and \ref{analysis}).  First
order changes in atmospheric extinction are removed through
differential (spectro)photometry, but second order colour-dependent
effects may become a significant systematic error for wide bands
and/or in the optical (see Bailer-Jones \& Lamm \cite{bjl} and
references therein).

The wavelength regions 1.35--1.45\,\micron\ and 1.80-2.00\,\micron\ --
which are generally inaccessible from the ground due to almost total
telluric absorption -- would be very useful bands (Fig. \ref{var5}),
because they trace the variation in the dominant water absorption
in UCD atmospheres.  Denoting these by t1 and t2 respectively, we see
that j1 is anticorrelated with both t1 and t2 with amplitudes of 0.04
and 0.05\,mags respectively, and j1$-$t1 and t1$-$t2 have amplitudes
of 0.11 and 0.12\,mags respectively, from a cond cloud on a dusty
atmosphere. This is larger than the best otherwise available j1, k2
signature.

\subsection{Spot and cloud signatures at 1700\,K and 2100\,K}

\begin{figure}[t]
\centerline{
\psfig{figure=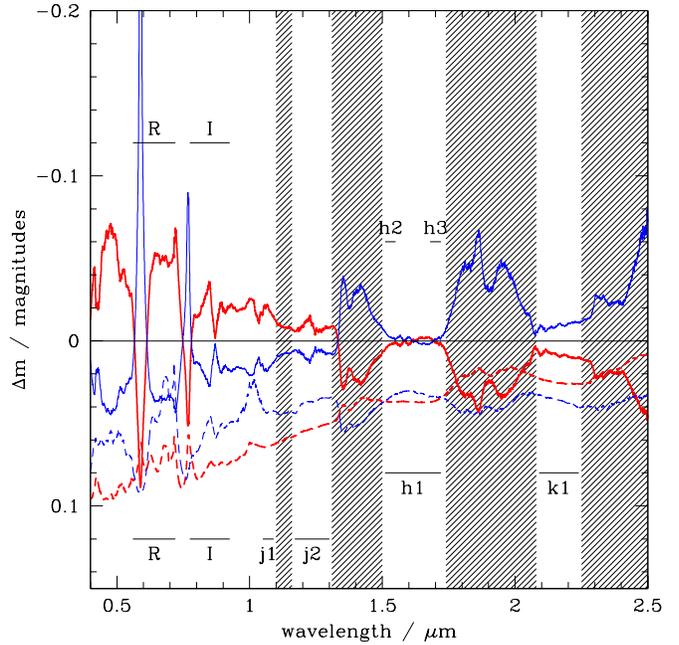,angle=0,width=0.5\textwidth}
}
\caption{Same as Fig. \ref{var5} but for a atmosphere with \teff\,=\,2100\,K.
[Colour]}
\label{var3}
\end{figure}

\begin{figure}[t]
\centerline{
\psfig{figure=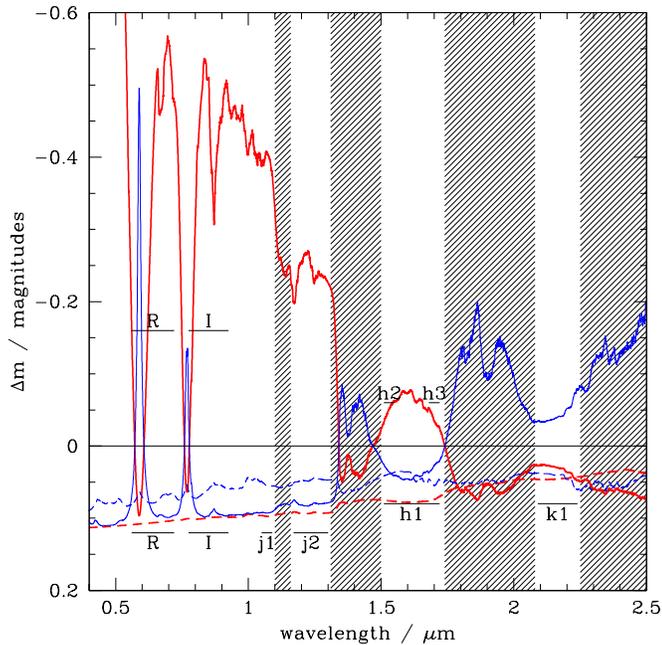,angle=0,width=0.5\textwidth}
}
\caption{Same as Fig. \ref{var5} but for a atmosphere with \teff\,=\,1700\,K.
Note the different scale on the vertical axis. [Colour]}
\label{var4}
\end{figure}

The above discussion is relevant for a UCD with an effective
temperature of 1900\,K and \logg\,=\,5.5. From the plots in Allard et
al.\ (\cite{allard01}), it appears that a different surface gravity will
make no detectable difference to the variability amplitudes
discussed. However, increasing or decreasing the effective temperature
by 200\,K has a significant effect, as can be seen in Figs. \ref{var3}
and \ref{var4}.

At 2100\,K, the cloud variability amplitude is reduced because of the
smaller differences between the opacity sources in the dusty and cond
models at higher temperatures, particularly blueward of
1.3\,\micron. This is simply because less dust condenses at higher
temperatures, so it makes less difference whether that
stays in suspension in the atmosphere or is precipitated out.  The
differences are enhanced at the lower effective temperature (1700\,K)
for the converse reason, i.e.\ more condensation of gas-phase species (and
their precipitation in the cond models).

The same bands as described above for 1900\,K can be used at 2100\,K,
but now the cloud signatures are reduced by a factor of four in the j1
and j2 bands. For a 10\% cloud coverage, this is probably below the
level at which systematic errors can be controlled from ground-based
observations.  However, at 2100\,K, the amplitude of the cool spot
signature is still relatively large, and amenable to detection.

At 1700\,K, the cloud signatures are greatly enhanced and are much
larger than the cool spot signatures, such that the j1$-$h1 colour has
an amplitude of 0.4\,mags due to a cloud and only 0.02\,mags due to a
cool spot (both covering 10\% of the surface). However, it must be
recalled that below 1800\,K UCDs are generally better fit by cond
atmospheres, and dusty clouds on such atmospheres show a much weaker
signature at 1700\,K (Fig. \ref{var4}). Moreover, the cloud signature
is probably sensitive to the {\it degree} of dust precipitation at these
intermediate temperatures and this is not accounted for in the present
models (see sections \ref{intro} and \ref{models}).

\subsection{Signature in the optical and mid infrared regions}
\label{optical}

I have focused on signatures in the NIR region as that is the region
observed and analysed in the rest of this paper. However, there are
strong signatures in the optical spectrum, as can be seen in Figs.\
\ref{var5}--\ref{var4}. There is a trade-off here, of course, because
ultra cool dwarfs are much fainter in the optical. The region
0.8--1.0\,\micron\ behaves in a similar fashion to the j1 band and the
region 0.65--0.72\,\micron\ shows an even stronger signature across a
range of effective temperatures. Very narrow bands centered on the
Na{\small I} (0.59\,\micron) and K{\small I} (0.77\,\micron) atomic
absorption lines are in principle a good discriminant between clouds
and cool spots when combined with a band in the range
0.8--1.1\,\micron, but the narrowness required ($\sim 0.02$\,\micron)
to avoid the opposite signature in the line wings may preclude their
use from a signal-to-noise ratio (SNR) point of view.

\begin{figure}[t]
\centerline{
\psfig{figure=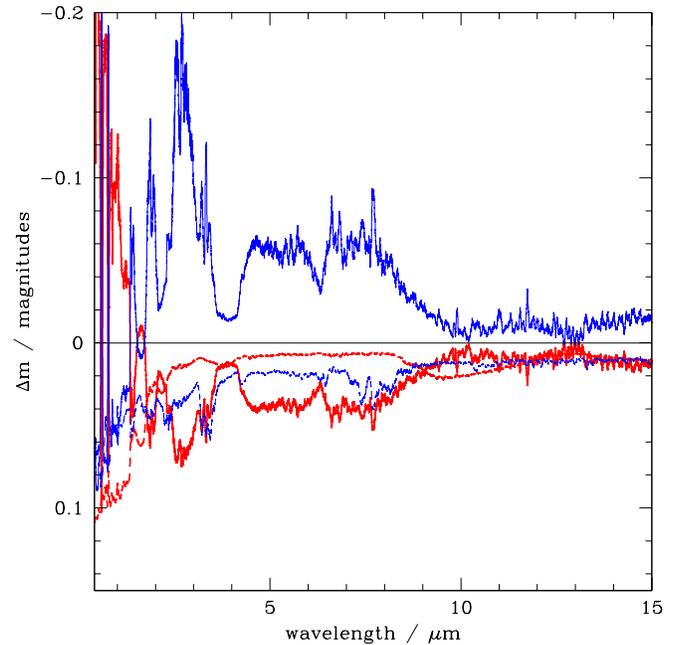,angle=0,width=0.5\textwidth}
}
\caption{Same as Fig. \ref{var5} but for the mid infrared wavelength region. [Colour]}
\label{var6}
\end{figure}

The mid infrared appears to contain no wavelength regions which are of
particular advantage over the near infrared for distinguishing cool
spots from dust-related clouds, at least not at around 1900\,K
(Fig. \ref{var6}). However, at lower temperatures methane forms, and
there remains the interesting possibility of looking for methane
clouds either in the near or mid infrared (Bailer-Jones, in
preparation).

\section{Data acquisition}
\label{data}

The analyses of the previous section were motivated by the desire
to interpret infrared spectra of the L1.5 dwarf 2M1145.  Spectra were
obtained using the Omega Cass infrared imager/spectrograph mounted on
the Cassegrain focus of the 3.5m telescope at Calar Alto, Spain.  The
detector is a 1024$\times$1024 HgCdTe HAWAII-1 focal plane array
sensitive between 0.8 and 2.5\,\micron.  The low resolution camera was
used, giving an image scale of 0.3$''$/pix. The resolution 420
($=\lambda/\Delta \lambda$) grism was used in combination with one of
two filters: With a JH filter it provided the wavelength range
0.98--1.84\,\micron\ unoverlapped in first order at a dispersion of
0.0021\,\micron/pix (plus a second order which was not used); with an
HK filter the wavelength range 1.48--2.41\micron\ was obtained in
first order at the same dispersion.  The slit had a width of 0.9$''$
and length of 141$''$.

The instrument was oriented such that the spectrograph slit was at an
angle $10.8^{\circ}$ East of North, to include both 2M1145 and the
bright reference star 79$''$ to the South (see section
\ref{objectives})..  The detector edges remained parallel to the slit.
The observing procedure was as follows.  Using a direct image, the
stars are brought to the approximate required position on the
detector. A guide star is acquired and the slit brought into the beam
at the exact position of the stars. Then the grism is inserted and
several sequences of observations made.  One {\it sequence} is defined
as two exposures (1,2) at the same telescope position, followed by a
telescope offset of 20$''$ along the slit and another two exposures
(3,4).  The same sequence was then normally repeated with the HK
filter. This was continued for between three and seven sequences in
each filter until it was noticed that the counts in the stars were
decreasing due to a decentering of the slit with respect to the stars
(which ranged from 0.6$''$ to 1.2$''$ over two hours, despite the
guiding). Then, without moving the telescope, grism or slit, the
telescope dome was moved across the field and a series of dome flat
fields obtained. Such a set of sequences constitutes a {\it
cycle}. This procedure was adopted in order to obtain flat fields with
as similar an optical path to the stars as possible and to keep this
path as constant as possible through minimal movement of the optical
elements. Each exposure used an integration time of 2\,min. The
readout time is essentially zero, so a single sequence took just over
8\,min.

Data were obtained on three consecutive nights starting on 29 February
2000 (Table \ref{obslog}). The seeing in JH was usually 1.0$''$--1.4$''$
but occasionally as high as 2.0$''$.  The weather was generally very
clear, but occasionally thin cirrus was present.

\begin{table}[t]
\caption[]{Observing log of spectrophotometric observations of 2M1145 in terms
of the number of {\it cycles} and {\it sequences}
in JH and HK (see text for definition of these terms).
Night 1 started near the end of 29 February 2000 UT at around JD2451604.4.}
\label{obslog}
\begin{tabular}{cccc}\hline
Night	& Cycle	& No.\ of JH	& No.\ of HK	\\
	&	& sequences	& sequences	\\
\hline
1	& 1	& 6		& 6		\\
	& 2	& 5		& 4		\\
	& 3	& 7		& 6		\\
	& 4	& 5		& 4		\\
2	& 1	& 7		& 7		\\
	& 2	& 5		& 5		\\
	& 3	& 3		& 3		\\
	& 4	& 7		& 6		\\
3	& 1	& 5		& 5		\\
	& 2	& 4		& 4		\\
\hline
\end{tabular}
\end{table}

\section{Image processing and spectral extraction}
\label{processing}

The data reduction and spectral extraction was mostly performed using
IRAF\footnote{the Image Reduction and Analysis Facility, provided by
the National Optical Astronomy Observatories.}.  Careful processing is
required to avoid biasing the data, so the procedure developed is
described.  The JH and HK spectra were treated separately.

From the four images (1,2,3,4) in a sequence, four {\it difference
images} were produced: a=$1-3$, b=$-$a, c=$2-4$, d=$-$c. This removes
most of the sky background plus the detector bias pattern.  These
images were then flat fielded using the end-of-cycle dome flats
described in section \ref{data}. These flats show absorption features
due to the air in the circa 50\,m optical path between lamp and
detector. However, as the dispersion function for the flats and the
science images is the same, this provides a valid flatfielding of the
detector in the spatial direction to put the two dither positions (a,c
and b,d) on the same photometric scale.  Small shifts of the slit with
respect to the detector in the wavelength direction change the flat
field value at each wavelength, but as this is the same change for
both 2M1145 and the reference star, it is not detrimental to
spectrophotometry in the {\it relative spectra} (that is, the ratio of
these -- see below).  Bad pixels (as identified from dark images) were
removed from the flat-fielded difference images by linearly
interpolating over them in the spatial direction.

These four images provide four independent spectra of each of the
target and reference stars. The spectra were extracted using the
one-dimensional optimal extraction technique implemented in IRAF's
`apsum' task. As these images are sky subtracted, pixel cleaning
(e.g.\ to remove cosmic rays) could not be used reliably within this
package (but the effect of cosmic rays was found to be minimal).  A
local sky subtraction at each wavelength using the median value of
regions on each side of the spectrum was found to be necessary to
further reduce the sky contribution. (The image differencing does not
remove changes in the sky background occuring over the four minutes
elapsing between images 1 and 3 or 2 and 4.)

The spectra were wavelength calibrated using 15--25 OH airglow lines
identified in Oliva \& Origlia (\cite{oliva92}) and present in spectra
extracted from the undifferenced images.  A quadratic dispersion
function was most suitable, giving RMS errors in the fits of less than
6e$^{-4}$\,\micron\ ($<0.3$ pixels). The spectra were then transformed
to a common linear wavelength scale with 0.0025\,\micron\ wide bins
using cubic spline interpolation. These spectra of the individual
objects shall be referred to as the {\it nominal spectra}. Theoretical
error spectra were calculated from these based on the Poisson noise
from star and sky and read noise contributions in the weighted
extraction aperture.

\begin{figure}[t]
\centerline{
\psfig{figure=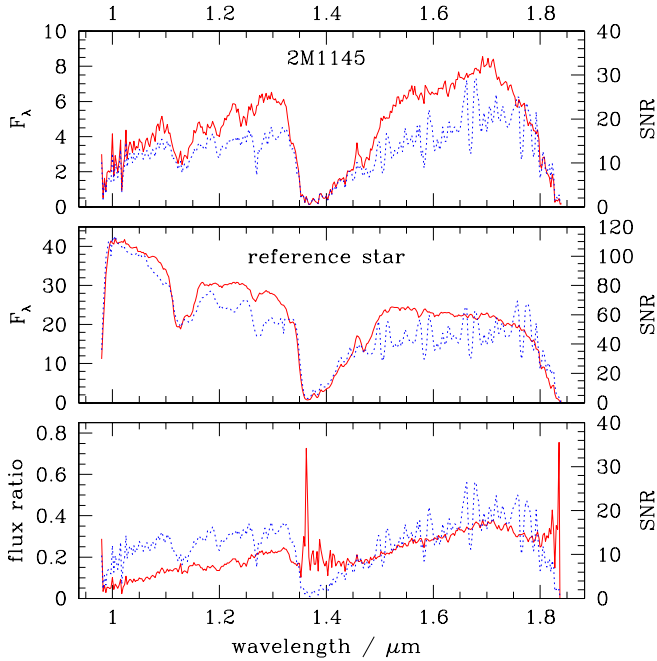,angle=0,width=0.5\textwidth}
}
\caption{Example spectra obtained in the JH filter. The top two panels
show the {\it nominal spectra} of 2M1145 and the reference star and
the bottom panel shows the {\it relative spectrum} (solid lines, left
axis). The corresponding signal-to-noise ratios (SNRs) are also shown
(dotted lines, right axis).  The nominal spectra shown are the
averages of four consecutive two minute exposures (i.e.\ from one {\it
sequence}).  The spectra of the two objects are obtained
simultaneously using a long slit.  A relative spectrum is formed by
taking the ratio of these two spectra. The combined relative spectrum
shown is the average of these four relative spectra (which is not
simply the ratio of the two nominal spectra shown). The spectra have
been flat fielded and sky subtracted, but not corrected for telluric
absorption.  Of course, many telluric features are absent from the
relative spectrum, as intended.  The flux scales for the two nominal
spectra are the same and in arbitrary units. The spectra are from the
fifth sequence on night 1. The low signal and SNR between 1.33 and
1.45\,\micron\ is the familiar opaque region of the Earth's atmosphere
between the H and K bands.}
\label{spec1}
\end{figure}

\begin{figure}[t]
\centerline{
\psfig{figure=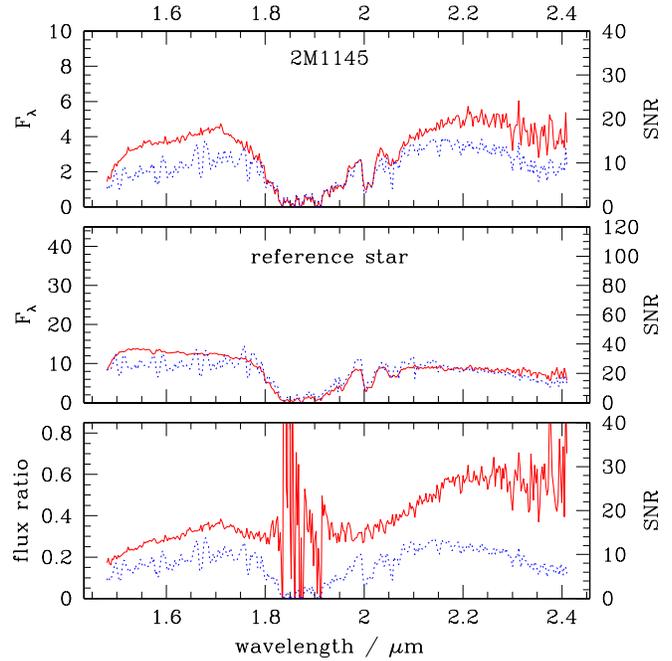,angle=0,width=0.5\textwidth}
}
\caption{Same as Fig.\ \ref{spec1} but for the HK filter. The spectra
are from the fifth sequence on night 1. The flux is in
the same (arbitrary) units as in Fig. \ref{spec1}.  The low signal and
SNR between  1.80 and 1.95\,\micron\ is due to strong water
absorption by the Earth's atmosphere.}
\label{spec2}
\end{figure}

{\it Relative spectra} are produced by dividing the target nominal
spectrum by the reference nominal spectrum from the same difference
image.  On the assumption that the reference star is photometrically
stable, this should remove any time-dependent Earth atmosphere
transmission variations from a time series of observations. As the
relative photometry is in a series of narrow wavelength bins
(0.0025\,\micron), second order colour-dependent extinction effects
are negligible. The four
relative spectra from a sequence are then averaged to create a higher
SNR combined spectrum.  This is shown in Figs.\
\ref{spec1} and \ref{spec2} along with the combined nominal spectra.
Some features are recognisable, such as the Na{\small I} line at
1.14\,\micron\ (at the bottom of a telluric water absorption band),
FeH at 1.20\,\micron\ and the resolved K{\small I} doublet at
1.25\,\micron.

A band magnitude, $m$ (see section \ref{bandsec}) is calculated by
\begin{equation}
m=-2.5\log_{10} \left( \frac{1}{4}\sum_{i=1}^{i=4}f_i \right)
\label{mageqn}
\end{equation}
where $f_i$ is the sum of the (relative) flux over the wavelength
interval for that band (Table \ref{bands}) for the $i^{th}$ (relative)
spectrum in that cycle (see section \ref{data}).  The idea is always
to {\it first} ratio the target measurement with the reference
measurement taken simultaneously and {\it then average}, thus removing
concurrent fluctuations in the Earth's atmospheric transmission.  The
same procedure is used for colours by replacing $f_i$ in equation
\ref{mageqn} with the ratio of fluxes obtained from a (relative)
spectrum. Colours formed in this way from relative spectra are
called {\it relative colours}.  As I only ever look at {\it changes}
in band magnitudes and colour indices, the zero point is
irrelevant. Note that the j1 and j2 bands cannot be obtained from the
HK data and the k1 band cannot be obtained from the JH data, so colour
indices between these cannot be formed from contemporaneous data. The
bands h1, h2 and h3 can be formed from either the JH or HK spectrum,
which I will distinguish where necessary using subscripts, e.g.\
h1\jh\ and h1\hk.

Photometric errors are calculated for the bands by propagation of
the errors in the nominal spectra. The validity of these theoretical
errors has been assessed by comparison to empirical errors obtained
from the standard error in the mean of the four relative spectra, and
it is found that they generally agree with a few tens of
percent. These empirical errors show some large fluctuations,
probably because they are obtained from the scatter between just four
spectra.  For this reason I use the theoretical errors throughout.

\section{Differential variability analysis}
\label{analysis}

\begin{figure}[t]
\centerline{
\psfig{figure=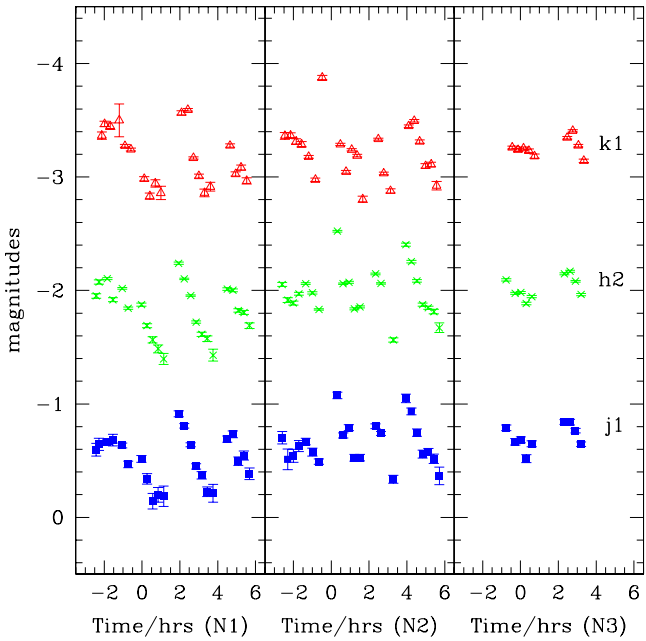,angle=0,width=0.5\textwidth}
}
\caption{Band magnitudes for the relative spectra of 2M1145 
(i.e.\ spectrum of 2M1145 divided by that of the reference star) for
j1 (solid square), h2\jh\ (cross) and k1 (open triangle). The
horizontal axis in each case is the time in hours from UT0 on nights
1, 2 and 3 (left to right). Each band has an arbitrary offset on the
vertical axis which is the same for all three nights. The bands are
defined in Table~\ref{bands}. [Colour]}
\label{res1}
\end{figure}

The first analysis was to see whether differential spectrophotometry
in the relative spectra is reliable. Fig.\ \ref{res1} shows this for
the three bands j1, h2\jh\ and k1. All three show a series of
correlated decreases in flux. The gaps between the successive trends
correspond to the transitions between observing cycles (section
\ref{data}). This effect is caused by the decentering of the star on
the spectrograph slit during a cycle. To affect the relative spectra,
the degree of decentering must be different for the two objects,
either because the slit did not remain parallel to the line joining
the stars, or because the slit edges were not sufficiently parallel to
each other.  Although such decentering was expected (from differential
flexing between the guide camera and the instrument), it was not
expected to be as large as 0.2\,mags per hour.  Thus we cannot perform
pure differential spectrophotometry with these data.

\begin{figure}[t]
\centerline{
\psfig{figure=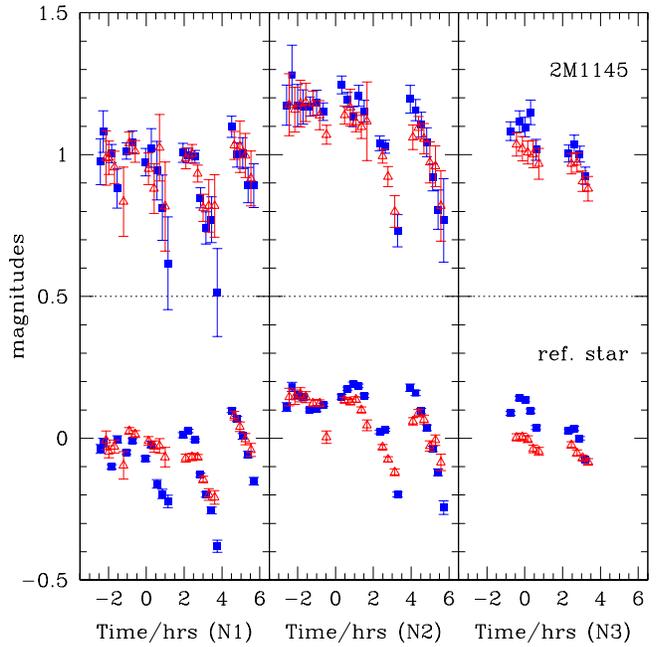,angle=0,width=0.5\textwidth}
}
\caption{Colour variability in the nominal (not relative) spectra
of 2M1145 and the reference star in the colours j1$-$h2 (filled
squares) and h2$-$k1 (open triangles). The colours have been
set to a mean (across all nights) of 0.0 for 2M1145 and 1.0
for the reference star. Points lying above
the dashed line are for 2M1145, those below are for the reference
star. The time axis is the same as in Fig. \ref{res1}. [Colour]}
\label{res2}
\end{figure}

Colour variability within 2M1145 alone also turns out to be affected
by the same problem.
Fig.\ \ref{res2} shows the variation in the colours
j1$-$h2 and h2$-$k1 formed from the nominal (rather than relative)
spectra of 2M1145 and the reference star. We immediately see that
within each cycle both objects appear to get bluer, i.e.\ more red light
than blue light is lost from the slit decentering. This is most likely
due to atmospheric refraction, such that the redder part of the refracted
image is that which is more decentered with respect to the slit.

\begin{figure}[t]
\centerline{
\psfig{figure=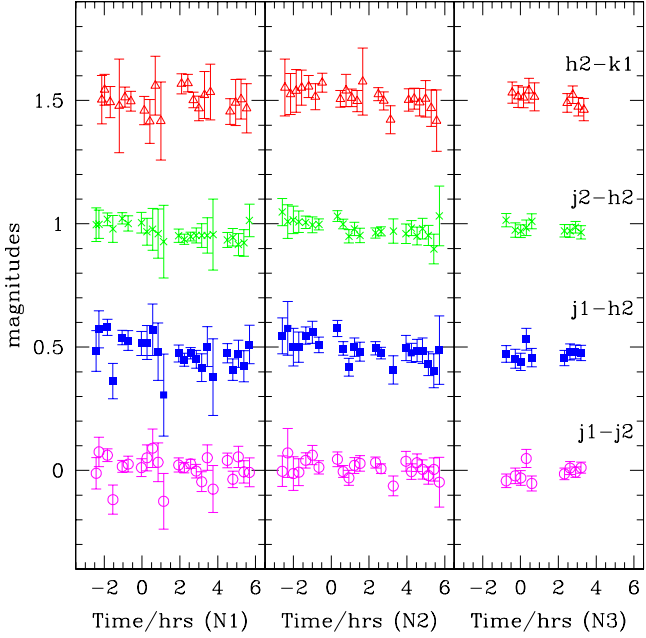,angle=0,width=0.5\textwidth}
}
\caption{Colours formed from the relative spectra (2M1145 divided by
the reference star). The colours shown are, from bottom to top,
j1$-$j2 (open circles), j1$-$h2 (filled squares), j2$-$h2 (crosses)
and h2$-$k1 (open triangles). Each colour has an arbitrary offset on
the vertical axis which is the same for all three nights. A more
positive magnitude means that 2M1145 gets redder relative to the
reference star.  The time axis is the same as in
Fig. \ref{res1}. Such light curves can, in principle, be used to
distinguish between variability due to spots and dust-related
clouds. [Colour]}
\label{res3}
\end{figure}

In spite of these problems, we find that {\it colour variability in
the relative spectra is reliable}, in the sense that there is no
systematic trend within a cycle. This can be seen in Fig. \ref{res3}
for the colours j1$-$j2, j1$-$h2, j2$-$h2 and h2$-$k1. That the
colours formed from the relative spectra can be reliable despite the
problems seen in Figs.\ \ref{res1} and \ref{res2} can be understood
from the fact that both 2M1145 and the reference star show the same
degree of `bluening' as a result of the slit decentering (see
Fig. \ref{res2}).

It is fairly clear from a visual inspection of Fig.\ \ref{res3} that
none of the four light curves shows good evidence for
variability. This can be put on a quantitative basis using a
statistical test, for example the \chisq\ test (see section 4.1 of
BJM). The probabilities that there is a real variability (i.e.\ not
related to the noise) are shown in Table \ref{amps}. This table also
gives the amplitudes of the relative colour light curves. It is
tempting to study the ratios of these amplitudes and compare them with
the predictions given in section \ref{predictions}. However, the
dominant contribution to these amplitudes is noise, so we would first
have to statistically subtract the noise contribution. This can only
be done once we have a model for both the noise and the underlying
variability. Taking the former to be Gaussian with standard deviation
$\sigma_{\rm noise}$ and the latter to be a sinusoidal signal, then
the RMS amplitude of the underlying signal, $\sigma_{\rm var}$
($=0.5\sqrt\pi$ times the peak-to-peak amplitude), is\footnote{This
can only be a rough approximation, however, as we do not expect
features which evolve faster than a rotation period to necessarily
show a sinusoidal signature. There is also the issue of how to form a
suitable single value of $\sigma_{\rm noise}$ given the large
variation in the errors in Fig. \ref{res3}.}
$\sqrt{\sigma^2_m-\sigma^2_{\rm noise}}$.  However, as we have no
evidence for significant variability in any {\it single} colour, this
is not appropriate for the present data set.

\begin{table}[t]
\caption[]{Amplitudes of the variations in the four relative colour
light curves shown in Fig.\ \ref{res3}.  Two measures of light curve
amplitude are given: \rms, the RMS (root-mean-square) of the relative
magnitudes, and \absdev, the average modulus deviation in the light
curve from its mean value (the linear equivalent of an
RMS). $\sigma_{\rm noise}$ is the median of the theoretical errors
(the error bars shown in Fig. \ref{res3}). $1-p$ is the probability
calculated from the \chisq\ test that the variability across all
points is ``real'', i.e.\ not due to the empirical photometric
errors.}
\label{amps}
\begin{tabular}{rrrrrrcc}
\hline
Colour	& \rms	& \absdev	& $\sigma_{\rm noise}$  & $1-p$	\\
	&  mags	& mags		& mags	                &         \\
\hline
j1$-$j2	& 0.044	& 0.033		& 0.034                 &  0.41  \\
j1$-$h2	& 0.055	& 0.040		& 0.044                 &  0.73  \\
j2$-$h2	& 0.032	& 0.026		& 0.032                 &  0.18  \\
h2$-$k1	& 0.040	& 0.031		& 0.054                 &  0.00  \\
\hline
\end{tabular}
\end{table}

\begin{figure}[t]
\centerline{
\psfig{figure=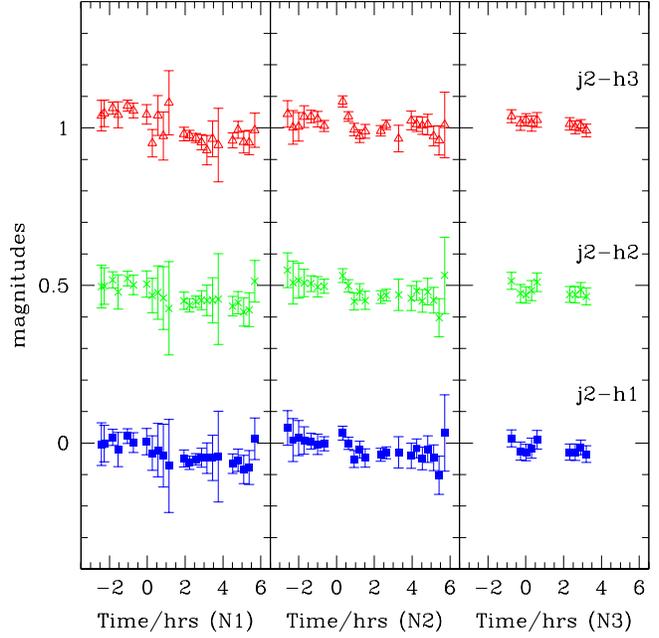,angle=0,width=0.5\textwidth}
}
\caption{As for Fig.\ \ref{res3} but for the relative colours (from
bottom to top) j2$-$h1 (filled squares), j2$-$h2 (crosses) and j2$-$h3
(open triangles). This figure demonstrates that the relative colour
light curves are essentially identical for all three H bands defined
in Table \ref{bands}. [Colour]}
\label{res4}
\end{figure}

The light curves in Fig.\ \ref{res3} are the same (within the errors)
if h1 or h3 is used instead of h2. This can be see in Fig. \ref{res4},
where the colours j2$-$h1, j2$-$h2 and j2$-$h3 have virtually the same
light curves.

\begin{figure}[t]
\centerline{
\psfig{figure=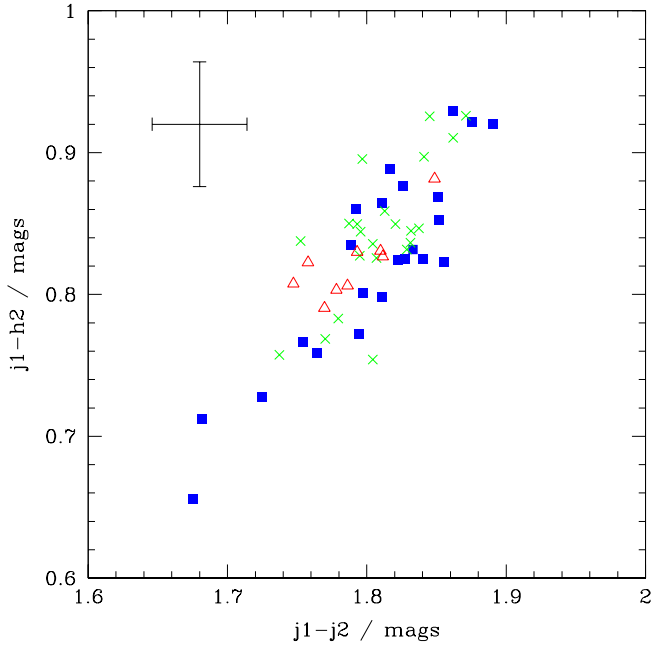,angle=0,width=0.5\textwidth}
}
\caption{Correlation between the two relative colours j1$-$j2 and j1$-$h2. The
different symbols refer to observations on the different nights: night
1 (solid squares); night 2 (crosses); night 3 (open
triangles). The median error is shown in the top-left corner. [Colour]}
\label{cor1}
\end{figure}

\begin{figure}[t]
\centerline{
\psfig{figure=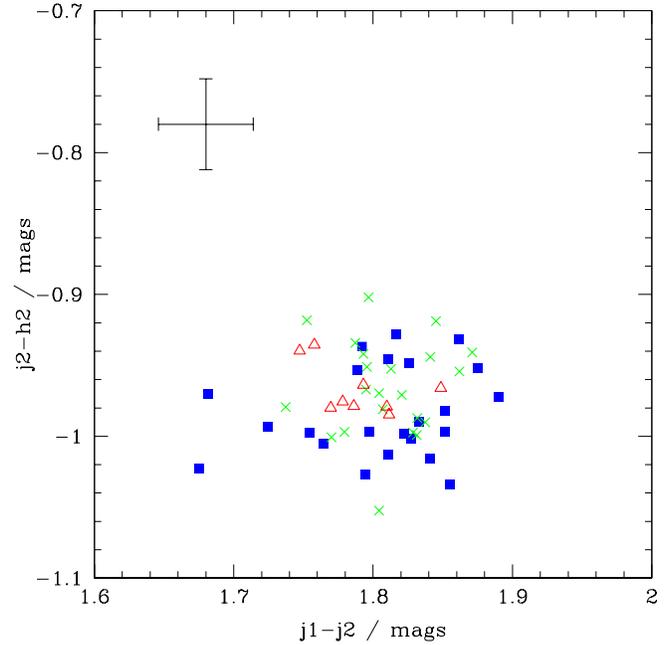,angle=0,width=0.5\textwidth}
}
\caption{Same as Fig. \ref{cor1} but for the colours j1$-$j2 and j2$-$h2. [Colour]}
\label{cor2}
\end{figure}

The \chisq\ test does not take into account any correlations between
the colour variations, which is what the observing procedure
was designed to exploit.  Fig.\ \ref{cor1} shows that there is in fact
a correlation in the relative colour variability for the colours
j1$-$j2 and j1$-$h2. Despite the large errors, there is a significant
positive correlation: the correlation coefficients for nights 1, 2 and
3 are 0.89, 0.71 and 0.81 respectively, and 0.83 for the whole data
set together. Random variations due to noise are very unlikely to give
such large correlations. There is also no trend of increasing or
decreasing colour within a night or an observing cycle (Fig.\
\ref{res3}), so no evidence that this colour correlation is telluric
in origin or an observational artifact. Thus while the noise on
individual measurements is too large to measure a significant
amplitude in a single colour, the colour variations are positively
correlated and indicate something intrinsic to the object observed
(either 2M1145 or the reference star). In contrast, Fig.\ \ref{cor2}
shows that j2$-$h2 has a negligible correlation with j1$-$j2 (the
correlation coefficients for nights 1, 2 and 3 are 0.20, 0.04 and 0.52
respectively, and 0.08 for all nights together).

As discussed in section \ref{predictions}, a positive correlation
between any of these colours alone in insufficient to distinguish
between cond clouds and cool spots on a dusty atmosphere.  However,
for a 200\,K cooler spot on a dusty atmosphere at 1900\,K or lower,
there is very little variation in j1$-$j2 compared to j1$-$h2 or
j2$-$h2. Therefore, in the presence of noise, we would expect both
j1$-$h2 and j2$-$h2 to be equally uncorrelated with j1$-$j2 for such a
spot.  We find that the former is correlated whereas the latter is
not. This is inconsistent with this particular cool spot model. For a
cond cloud we would expect both to be correlated, but in the presence
of noise we would observe a larger correlation between j1$-$h2 and
j1$-$j2 than between j2$-$h2 and j1$-$j2 simply because the amplitude
is larger. This difference becomes more pronounced at lower
temperatures.  Thus the observed correlations are more consistent with
a cond cloud than a cool spot, for a dusty atmosphere at 1900\,K.
Note, however, that a cool spot on a hotter dusty atmosphere (2100\,K,
see Fig.\ \ref{var3}) would produce a similar effect to that
seen. Thus until we can measure amplitudes more precisely (through
higher precision observations and an appropriate statistical
subtraction of the errors), the {\it interpretation} of this colour
correlation should be treated with some caution.

\begin{figure}[t]
\centerline{
\psfig{figure=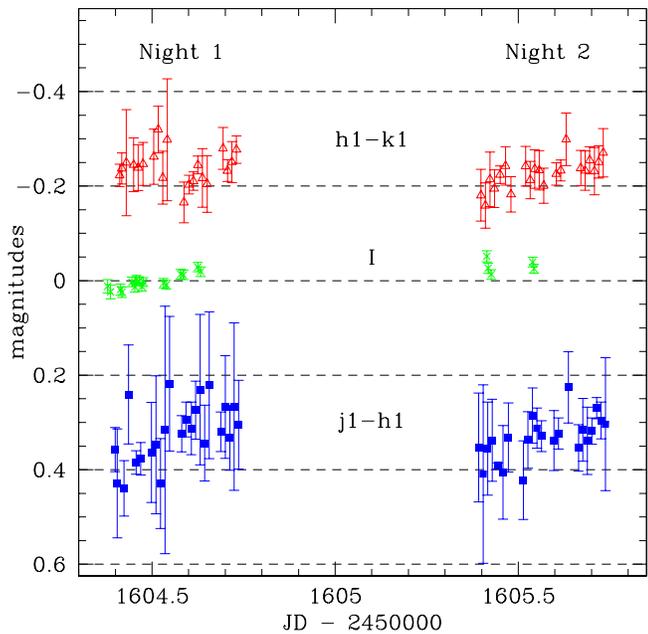,angle=0,width=0.5\textwidth}
}
\caption{Variation in 2M1145 of the relative colours j1$-$h1 (filled
squares) and h1$-$k1 (open triangles) plotted along with the
differential I-band photometry of 2M1145 (crosses) obtained
simultaneously on the 2.2m telescope at the same site (latter taken
from Fig.\ 3 of BJM).  Moving up the page the I-band gets brighter
and the colours {\it bluer}. The night labels refer to the present
data set (Table \ref{obslog}). Note that Fig.\ 3 of BJM is plotted the
other way up (see that paper's erratum). [Colour]}
\label{iband}
\end{figure}

The data in the present paper were obtained quasi-simultaneously with
the I-band data reported in BJM. The latter are reproduced in Fig.\
\ref{iband} for the overlapping time interval (correcting for the axis
errors in BJM reported in the erratum). Alongside this is shown the
variability in two relative colours from the present data set.  On
night 1, there is reasonable evidence to suggest that as 2M1145 became
brighter in I it also became bluer in j1$-$h1. Both changes are
significant according to the \chisq\ test ($1-p=0.99$ for I,
$1-p>0.99$ for j1$-$h1). This is still significant if h1 is replaced
with h3, but not so significant ($1-p=0.90$) for j1$-$h2.  This is the
signature expected for the appearance of a cond cloud on a dusty
atmosphere or the dissipation of a cool spot on the same atmosphere
(Fig.\ \ref{var5}).
  
\section{Summary and Discussion}
\label{discussion}

To summarise the results of the analysis for 2M1145: a formal \chisq\
test does not reveal any variability at the 99\% confidence level
across the whole data set in any of the four relative colour indices
given in Table \ref{amps}. Relative colours means the ratio between
two band fluxes, each formed by summing flux in a relative spectrum
(2M1145 divided by reference) over a specific wavelength region (see
section \ref{processing}).  Relative photometry and absolute colours
(i.e.\ the colour of 2M1145 alone) could not be used due to
time-dependent light losses from the slit. The relative colours appear
unaffected by this and there is no reason to consider them as
unreliable. Although no {\it single} colour shows significant
variation (across the whole observing period), there is a significant
correlation between the j1$-$j2 and j1$-$h2 relative colours. This is
not expected from random noise and does not appear to be caused by the
slit losses. I therefore conclude it to be intrinsic to either 2M1145
or the reference star. The reference star is most likely to be mid or
late-type F dwarf and not a short period variable (see section
\ref{objectives}) so the variability is probably not in this
object. Assuming that the variability is intrinsic to 2M1145, then it
is more consistent with dust-free (cond) clouds rather than a dusty
cool spot, assuming 2M1145 to have an overall dusty atmosphere and an
effective temperature of 1900\,K. However, if 2M1145 is actually
hotter (2100\,K), then cool spots are equally consistent with these
data. It was found that the contemporaneous I-band data of 2M1145 from
BJM show a slight correlation with the present data in the sense that
2M1145 became brighter in I as it became bluer in j1$-$h1. Although
the data were obtained at the same site, BJM have argued that the
I-band variability is not due to changes in the Earth's atmosphere, so
we can consider this correlation to be real. It is consistent with
either a cool spot or a cond cloud on a dusty atmosphere.

The present data allow upper limits to be placed on the
size of surface features
based on what peak-to-peak amplitude would have been required to
produce a significant \chisq\ ($1-p>0.99$). Assuming a constant noise
RMS of $\sigma_{\rm noise}$, variability could be detected in a
relative colour if the underlying (i.e.\ noise-free) {\it
peak-to-peak} amplitude were greater than about 1.4$\sigma_{\rm
noise}$.  Using the median noise in each relative colour from Table
\ref{amps}, this can be compared with the signatures of the spots
described in section \ref{predictions}.

If 2M1145 has a dusty atmosphere and an effective temperature of
1900\,K, then the data require that any coherent cond cloud features
cover less than 10--15\% of the projected surface. Alternatively, cool
spots (with $\Delta T=$\,200\,K) must cover less than 20\% of the
surface.  Even cooler spots would be restricted to a smaller coverage.
If a cond atmosphere is assumed instead of a dusty one, then dusty
clouds on this have essentially the same signature as cond clouds on
the dusty atmosphere (see Fig.\ \ref{var5}).  Cool spots, on the other
hand, have a weaker signature and could cover up to about 40\% of the
surface before being detected.

If 2M1145 is warmer (with a dusty atmosphere), then clouds could be
larger: At 2100\,K, a cloud coverage of up to 50\% would be possible
before it is detected by the measurements of this sensitivity. The
limits on cool spots are not changed at this temperature (see Fig.\
\ref{var3}).  If 2M1145 is cooler (1700\,K), then clouds are limited
to being much smaller: a coverage of just 2\% would have given rise to
a detectable signature (Fig.\ \ref{var4}). Large cool spots, on the
other hand, covering up to 40\% of the surface, are permitted. At this
lower temperature the dusty models may not be valid (due to increased
dust precipitation), so cond models should be considered. For a cond
atmosphere at 1700\,K, cool spots can still cover a large fraction of
the surface (40--60\%) before giving a detectable signature in these
bands, but note that these bands were not designed with such a
situation in mind. Dusty clouds are restricted to a coverage of no
more than 20\%. In all cases, these limits apply to a single
cloud/spot or several clouds/spots behaving coherently. Clearly, many
small features evolving independently of each other will have no net
effect on the light curve of the unresolved disk.

These limits can be compared with the feature sizes derivable from the
detections of variability in 2M1145 by BJM, specifically the derived
amplitudes and noises given in their Table 2. Using the noise
subtraction method in section \ref{analysis} of the present paper, the
99-01 and 00-02 data from BJM correspond to peak-to-peak signal
amplitudes of 0.025 and 0.018\,mags respectively. For a dusty
atmosphere at 1900\,K, these translate to clear cloud coverages of
2.5\% and 2\% respectively, and about the same for 200\,K cooler
spots.  These are well below the upper limits imposed by the present
infrared data and therefore entirely consistent with them.

All of these limits are based on the assumptions (1) that the limiting
cases of the AMES-dusty and AMES-cond atmospheres of Allard et al.\
(2001) are appropriate, (2) that 2M1145 was not observed during an
abnormal phase of atmospheric or magnetic inactivity, and (3) that the
timescale for the evolution of surface features in UCDs is much less
than the 54 hour observing period. The case for the latter was argued
by BJM.

Clearly, the present work can be improved upon in a number of ways,
not least through having a more stable instrument to avoid slit
losses. A wider slit may nonetheless be required to achieve pure
relative spectrophotometry in order to accommodate brightness changes
due to seeing fluctuations.  This may require observing in the optical
where the sky background is reduced and there are still significant
features (see Fig.\ \ref{var5} and section \ref{optical}). While L
dwarfs are much fainter here (Fig.\ \ref{sed1}), this may still be a
more effective method if the data quality is limited by systematic
errors and not photon noise.  Alternatively, J and K band {\it
imaging} has a significant signature, and has the advantage that many
reference stars can be used to calibrate sky variations (see
Bailer-Jones \& Lamm \cite{bjl}). Ultimately, Doppler imaging of the
brightest, most rapidly rotating UCDs provides the possibility to
monitor surface features in a more direct manner.

\section{Conclusions}
\label{conclusions}

Non-periodic photometric variability in late M and L dwarfs has been
indicated by a few observing programs to date. In this paper, I
have presented the results of a program to spectrophometrically
monitor the known variable L1.5 field dwarf 2M1145 in the near
infrared (1--2.5\,\micron), in an attempt to identify the cause of its
variability. Primary candidates are the temporal evolution of
cool magnetically-induced spots and dust-related photospheric
clouds. Using the spectral models of Allard et al.\ (\cite{allard01}),
I have demonstrated the photometric signatures of these two types of
variability in various bands across the near infrared spectrum. For
the evolution of either a clear, dust-precipitated cloud against a
dusty atmosphere or a dusty cloud  against a dust-cleared atmosphere,
there is an anticorrelated variability in the J and K bands. At an
effective temperature of 1900\,K, the peak-to-peak amplitudes are 0.07
and 0.03 magnitudes respectively for a cloud covering 10\% of the
surface, both decreasing by a factor of about three at 2100\,K, and
the J band signature increasing by a factor of over five at 1700\,K. A
cool spot, on the other hand, shows a correlated J,K signature at all
temperatures.

The observational data, which cover several hours on three consecutive
nights, show no evidence for variability in any of four individual
colour indices. However, there is evidence for {\it correlated
variations} in the colours which cannot be attributed to noise or
artifacts of the observing method. These changes are more consistent
with the presence of a dust-cleared cloud on a dusty atmosphere than a
cool dusty spot on such an atmosphere. The non-detections in the
individual bands permit upper limits to be placed on the size of any
clouds or spots.  Adopting a dusty atmosphere for 2M1145 with an
effective temperature of 1900\,K, the data restrict clear clouds to
cover less than 10--15\% of the surface, or a 200\,K cooler spot (or
spots) to cover less than 20\% of the surface.  If 2M1145 is as cool
as 1700\,K and still has a dusty atmosphere, then clear clouds are
restricted to cover less than 2\% of the surface.  However, if a
dust-cleared model is more appropriate at 1700\,K, the upper limit on
cloud coverage is more like 20\%.

\begin{acknowledgements}
I am grateful to Frank Valdes for advice on some of the more subtle
aspects of spectral extraction with IRAF, and to Christiane Helling
and Peter Woitke for stimulating conversations on the dynamics of dust
in (sub)stellar atmospheres.  I would also like to thank France
Allard, Peter Hauschildt and Reinhard Mundt for useful comments on a
draft of this paper.
\end{acknowledgements}

\end{document}